\documentclass[showpacs,aps,pra,twocolumn]{revtex4}
\usepackage{amsmath}
\usepackage{epsfig}
\usepackage{graphicx}
\usepackage{color}
\usepackage{amssymb}
\usepackage{epstopdf}
\usepackage{array}
\usepackage{tabularx}
\usepackage[utf8]{inputenc}
\usepackage[T1]{fontenc}
\usepackage{float}
\usepackage{bbold}
\usepackage[normalem]{ulem}
\usepackage{dsfont}
\begin{document}
\title{Effective potentials in a rotating spin-orbit-coupled spin-1 spinor condensate}
\author{Paramjeet Banger$^{1}$\footnote{2018phz0003@iitrpr.ac.in}}
\author{R. Kishor Kumar$^{2}$\footnote{kishor.ramavarmaraja@otago.ac.nz}}
\author{Arko Roy$^{3}$\footnote{arko@iitmandi.ac.in}}
\author{Sandeep Gautam$^{1}$\footnote{sandeep@iitrpr.ac.in}}
\affiliation{$^{1}$Department of Physics, Indian Institute of Technology Ropar, Rupnagar  140001, Punjab, India}
\affiliation{$^{2}$Department of Physics, Centre for Quantum Science, and Dodd-Walls Centre for Photonic and\\ Quantum Technologies, University of Otago, Dunedin 9054, New Zealand}
\affiliation{$^{3}$School of Physical Sciences, Indian Institute of Technology Mandi, Mandi-175075 (H.P.), India}
\begin{abstract}

We theoretically study the stationary-state vortex lattice configurations of rotating spin-orbit- and
coherently-coupled spin-1 Bose-Einstein condensates trapped in quasi-two-dimensional harmonic potentials. 
The combined effects of rotation, spin-orbit and coherent couplings are analyzed systematically 
from the single-particle perspective. Through the single-particle Hamiltonian, which is exactly solvable 
for one-dimensional coupling, under specific coupling and rotation strengths, we illustrate that a boson 
in these rotating spin-orbit- and coherently-coupled condensates are subjected to effective toroidal, 
symmetric double-well, or asymmetric double-well potentials. 
In the presence of mean-field interactions, using the coupled Gross-Pitaevskii formalism at moderate to high rotation 
frequencies, the analytically obtained effective potential minima and the numerically obtained coarse-grained density 
maxima position are in excellent agreement. On rapid rotation, we further find that the 
spin-expectation per particle of an antiferromagnetic spin-1 Bose-Einstein condensate approaches unity 
indicating a similarity in the response with ferromagnetic spin-orbit-coupled condensates.

\end{abstract}
\maketitle
\section{Introduction}
\label{Sec-I}
The experimental realization of artificial gauge fields \cite{lin1, lin2} and
spin-orbit (SO) coupling between the spin and the linear momentum
of electrically neutral bosons \cite{soc-first} opened a hitherto inaccessible 
research direction to the researchers.
An SO coupling, which couples the three magnetic sublevels of spin-1 $^{87}$Rb, wherein Rashba \cite{Rashba} and 
Dresselhaus \cite{Dresselhaus} couplings contribute with equal weights has been realized 
in an experiment \cite{Campbell2016}. 
More recently, using an optical Raman lattice, two-dimensional SO coupling and the resulting topological 
bands have been experimentally realised with $^{87}$Rb involving two of its hyperfine spin states \cite{wu}.
In spin-1 Bose-Einstein condensates (BECs), SO coupling results in
diverse ground-state phases like plane-wave, stripe or standing-wave, vortex-lattice, zero-momentum phases,
etc. \cite{plane-stripe,Ruokokoski2012-rapidcommu,stringari-spin1,zhai2015degenerate}. Besides these unusual
phases, self-trapped vortex solitons \cite{sandeep-vortexsoliton1,sandeep-vortexsoliton2}, knotted solitons \cite{knotted-soliton},
super-stripes, and super-lattices \cite{adhikari} can also emerge as the ground state solutions of the SO-coupled spin-1
BECs in different parameter domains. Furthermore, coupling between the spin and orbital angular momentum 
of a neutral bosonic atom has also been experimentally realized recently \cite{lin3, lin4}.

The realization of the SO coupling has made it possible to explore the interplay of synthetic non-Abelian and Abelian gauge potential 
arising due to rotation ~\cite{jeandalibard_mod}, e.g. in the two-component SO-~\cite{radic,xu,zhou,liu1,aftalion,Shi} or
coherently-coupled pseudospin-1/2 BECs~\cite{Aftalion_spin_half}.
In spin-1 BECs, the interaction between Rashba 
SO coupling and rotation under rapid quenching leading to half-skyrmion excitations~\cite{liu2}, 
hexagonal lattice of skyrmions and a square 
lattice of half-quantized vortices \cite{Su} has already been theoretically investigated.
It has also been shown numerically that a rotating spin-1 BEC with anisotropic SO coupling can support 
vortex-chain solutions \cite{liu3,rotating_spin1-aniso}, whereas the presence of an isotropic  Rashba SO coupling 
may result in a vortex lattice with a hexagonal or an approximate square lattice pattern \cite{vortex_lattice-adhikari}.
The numerical studies on ground states of rotating Rashba SO-coupled gases in concentrically 
coupled toroidal traps \cite{necklace_torroidal}, rotating ferromagnetic BEC with isotropic 
three-dimensional SO coupling \cite{weyl_soc}, and SU(3) coupling subjected to 
a magnetic-field gradient \cite{su3} are among the other investigations 
which deserve to be mentioned.
More recently, topological vortical phase transitions in an SO-coupled spin-2 BEC under 
rotation have been theoretically studied \cite{Zhu}.

In this paper, we highlight the combined effects of spin-exchange interactions, SO and coherent couplings, and 
rotation frequency on spin-1 condensates.
We consider a quasi-two-dimensional harmonically trapped spinor
condensate with a generic Rashba SO coupling of the form $\propto (\gamma_x S_x\partial/\partial x + \gamma_y S_y\partial/\partial y)$ \cite{Rashba} and a coherent coupling. 
Here $S_x$ and $S_y$ are the spin operators for the spin-1 system and $\gamma_x$ and $\gamma_y$ are the SO-coupling strengths. We proceed to study the stationary-state solutions with a focus primarily on 
moderate to large rotation frequencies of up to $0.95$ times of the trapping frequency and within the domain of validity of the 
mean-field model.  It is important to point out that the celebrated 
experimental realization of SO coupling \cite{soc-first} corresponds to one dimensional coupling 
$\gamma_xS_x p_x$ with {a} non-zero coherent coupling. To execute our studies, 
from a single-particle perspective, we examine in detail the effective potential 
arising out of rotation, SO, and coherent couplings and relate it
to the scalar and vector potentials experienced by a boson for an experimentally relevant case. 
We compute the spin-expectation per particle as a function of rotation frequency and 
one of the key findings of the work is a similarity 
in the response of SO-coupled $^{87}$Rb (ferromagnetic) and $^{23}$Na (aniferromagnetic) BECs at moderate 
to large rotations; we also examine the spin-texture at small 
and large rotation frequencies to buttress this point. The inclusion of one-dimensional
SO coupling, i.e., $\gamma_x\ne 0,\gamma_y=0$, 
coherent coupling, and a theoretical analysis of an SO-coupled single-particle Hamiltonian
leading to the evaluation of effective potentials makes this study distinct from earlier studies on
SO-coupled BECs under rotation \cite{xu,vortex_lattice-adhikari}. 


The paper is organized as follows.
The analytic solutions of the single-particle Hamiltonian corresponding to the SO- and coherently-coupled 
bosons under rotation are provided in Sec.~\ref{Section-II}. To describe the realistic experimental scenario, 
the mean-field density and spin-dependent interaction terms are included to formulate the coupled 
Gross-Pitaevskii equations (GPEs) in Sec.~\ref{Section-III}. The stationary-state solutions
for the interacting SO- and coherently-coupled $^{87}$Rb and $^{23}$Na spin-1 BECs in a rotating 
frame are then obtained and discussed in Sec.~\ref{Section-III-A}. The response of the system is 
further explored through the computation of the spin-expectation per particle as a function of rotation 
frequency and illustrated in Sec.~\ref{Section-III-B}. We conclude highlighting the main results in Sec.~\ref{Section-IV}.

\section{Single-particle Picture}
\label{Section-II}
Under a two-dimensional harmonic confinement, the Hamiltonian of an SO- and coherently-coupled spin-1 boson in the rotating frame in the 
dimensionless form is given by 
\cite{xu,necklace_torroidal}
\begin{align}\label{single_pa}
H_0 =& \left[\frac{p_x^2+p_y^2}{2}+V(x,y) - \Omega_{\rm rot} L_z\right]\mathds{1}\nonumber  \\ 
&+ \gamma_x S_x p_x+\gamma_y S_y p_y + \Omega_{\rm coh}S_x,
\end{align}
where $p_{\nu} = -i \partial/\partial_{\nu}$ with $\nu = (x, y)$, $V(x,y) = (x^2+y^2)/2$ is an isotropic harmonic 
trapping potential, $\Omega_{\rm rot}$ is the angular frequency of rotation around $z$ axis, 
$L_z =\left( xp_y- yp_x\right) \label{lz}$ is the $z$ component of the 
angular-momentum operator, $\mathds{1}$ is a $3\times 3$ identity matrix, $\gamma_x$ and $\gamma_y$ are 
the SO-coupling strengths, $\Omega_{\rm coh}$ is the coherent-coupling strength, and $S_\nu$ is the
irreducible representations of angular momentum operator for a spin-1 system. 
The units of length, time, energy, and 
energy eigenfunctions are considered to be 
$a_{\rm osc} = \sqrt{\hbar/(m\omega_x)}$, $\omega_x^{-1}$, $\hbar\omega_x$, and $a_{\rm osc}^{-1}$, 
respectively, where $m$ is the mass of the boson and $\omega_x$ is the harmonic oscillator frequency along 
$x$ direction.
To delineate the combined effect 
of rotation, SO, and coherent couplings, we calculate the minimum-energy 
eigenfunctions and eigenenergies  of the Hamiltonian for two analytically tractable cases: 
\begin{subequations}
\begin{align}
\quad \gamma_x & \ne0,~\gamma_y = 0,~\Omega_{\rm coh} \ne 0,\label{case-a}\\
\quad \gamma_x &= \gamma_y\ne 0,~\Omega_{\rm coh} = 0,\label{case-b}
\end{align}
\end{subequations}
where (2a) represents an experimentally realizable equal-strength mixture of Rashba and Dresselhaus couplings \cite{soc-first,Campbell2016}, 
which couples the spin with the linear momentum along $x$ direction, and 
the latter (2b) employs the Rashba SO coupling \cite{Rashba} which couples the spin with linear momentum along $xy$ plane.

The calculation of the eigen-spectrum of $H_0$ in the former case (\ref{case-a}) is facilitated by a unitary transformation, 
$H_0\rightarrow U^\dagger H_0U$ with
\begin{equation}\label{Udagger}
U=\begin{pmatrix}
1/2 & -1/\sqrt{2} & 1/2\\
1/\sqrt{2} & 0 & -1/\sqrt{2}\\
1/2 & 1/\sqrt{2} & 1/2
\end{pmatrix},
\end{equation}
as the rotation operator. The operator $U$ rotates the spin state about $y$ axis in an anticlockwise direction by an angle 
$\pi/2$ \cite{Cohen-Shankar}. The transformed Hamiltonian 
 $U^\dagger H_0U = {\rm diag} \left(h_{+1},h_0,h_{-1}\right) ,\label{decoupled_ev}$
where $\rm {diag}\left(\ldots\right)$ stands for a $3\times3$ diagonal matrix (operator). 
The operators $h_{j}$s  are
\begin{align}
h_j 
    =& \frac{(p_x + \frac{j \gamma_x}{1-\Omega_{\rm rot}^2}+\Omega_{\rm rot}\bar{y})^2}{2}+\frac{(p_{\bar{y}}-\Omega_{\rm rot}x)^2}{2}\nonumber\\ 
     & +(1-\Omega^2_{\rm rot})\frac{(x^2+\bar{y}^2)}{2} -\frac{j^2\gamma_x^2}{2(1-\Omega^2_{\rm rot})}+j \Omega_{\rm coh},\label{h_j}
\end{align}
where $j = 0,\pm 1$, $\bar{y}=y-j \gamma_x \Omega_{\rm rot}/(1-\Omega^2_{\rm rot})$ and
$p_{\bar y} = -i\partial /\partial \bar{y}$ is the canonical conjugate momentum of
${\bar y}$. The decoupled eigenvalue equation for $h_j$ is $h_j\psi_j(x,\bar{y}) = E_j \psi_j(x,\bar{y})$
which can be simplified by substituting $\psi_j(x,\bar{y}) = 
\bar{\psi}_j(x,\bar{y})\exp\left(-i\frac{j \gamma_x}{1-\Omega_{\rm rot}^2} x\right)$
to obtain
 \begin{eqnarray}
 \Big[\frac{p_x^2 +p_{\bar{y}}^2}{2} +\frac{x^2+\bar{y}^2}{2}-\Omega_{\rm rot}(x p_{\bar{ y}}&-&\bar{y}p_x)
  -\frac{j^2\gamma_x^2}{2(1-\Omega^2_{\rm rot} )}\nonumber\\
  +j \Omega_{\rm coh}\Big]\bar{\psi}_j(x,\bar{y})&=& E_j \bar{\psi}_j(x,\bar{y}).\label{iso_ham}
\end{eqnarray}
The Hamiltonian on the left hand side of Eq.~(\ref{iso_ham}), barring the constant terms, is that
of a two-dimensional isotropic harmonic oscillator under rotation. It is to be noted that the eigenfunctions of this Hamiltonian 
are also the eigenfunctions of the Hamiltonian representing a isotropic harmonic oscillator in the absence of rotation 
which commutes with $L_z$~\cite{Cohen-Shankar}.
The ground state eigenenergy is, therefore, given by
\begin{equation}
  E_{j} =\frac{2(1 +j\Omega_{\rm coh})(1-\Omega_{\rm rot}^2) -j^2\gamma_x^2}
                     {2 (1-\Omega_{\rm rot} ^2)},
\end{equation}
and the corresponding eigenstate is
$\bar{\psi}_j(x,\bar{y}) = \frac{1}{\sqrt{\pi}}\exp\left(-\frac{x^2+\bar{y}^2}{2}\right)$.
On the $x$-$y$ plane, we thus obtain
\begin{equation}\label{c1_ansatz_ana}
\psi_j(x,y) = \frac{1}{\sqrt{\pi}}\exp\left[-\frac{x^2+\left(y-\frac{j\gamma_x\Omega_{\rm rot}}{1-\Omega_{\rm rot}^2}\right)^2}{2}-i\frac{j\gamma_x}{1-\Omega_{\rm rot}^2} x\right].
\end{equation}
The three minimum-energy vector eigenfunctions of the original Hamiltonian $H_0$ with eigenenergies
$E_j$s can now simply be written as
$\Phi_j(x,y) = \psi_j(x,y) U \zeta_j$,
where  $U$ and $\psi_j(x,y)$ are defined in Eqs.~(\ref{Udagger}) and (\ref{c1_ansatz_ana}), respectively.
Here $\zeta_j$s are the three eigenvectors of $S_z$:
$\zeta_{+1} = \left(1,~0,~0\right)^T$, $\zeta_{0} = \left(0,~1,~0\right)^T$, and $\zeta_{-1}  = \left(0,~0,~1\right)^T$,
where $T$ denotes the transpose.
In the absence of coherent coupling, $\Omega_{\rm coh} = 0$, the eigenfunctions $\Phi_{-1}(x,y)$ and $\Phi_{+1}(x,y)$ become degenerate having the least energy. Under these considerations, the principle of linear superposition further admits $c_{+}\Phi_{+1} + c_{-}\Phi_{-1}$ to be a possible degenerate eigenfunction subject to the constraint $|c_{+}|^2+|c_{-}|^2=1$. In the
presence of infinitesimally small repulsive interactions, say spin-independent interactions, the interaction 
energy is minimized if $|c_+| = |c_{-}| = 1/\sqrt{2}$ resulting in a equal-strength mixture of 
$\Phi_{\pm 1}(x,y)$. The resultant density, $\left[|\psi_{+1}(x,y)|^2+|\psi_{+1}(x,y)|^2\right]/2$,  
is bimodal with equal-height peaks at $(0,\pm \gamma_x\Omega_{\rm rot}/(1-\Omega_{\rm rot}^2))$; 
this is indeed reflective of an effective two-well potential experienced by the boson. 
The presence of coherent coupling $\Omega_{\rm coh}\neq 0$, however, results in the 
lifting of the degeneracy between $\Phi_{-1}(x,y)$ and $\Phi_{+1}(x,y)$ with 
$\Delta E = E_{+1} - E_{-1} 
         = 2\Omega_{\rm coh}.$

The exact effective potential experienced by the boson can also be computed through
vector and scalar potentials \cite{radic}. To identify these potentials for the former case (2a), 
we rewrite $h_j$ in Eq.~(\ref{h_j}) as
\begin{equation}
h_j = \frac{(p_x - A_x^j)^2}{2} + \frac{(p_y - A_y^j)^2}{2} + W_j(x,y) + V(x,y),
\end{equation}
where $A_x^j = - j\gamma_x - \Omega_{\rm rot} y,\quad A_y^j = \Omega_{\rm rot} x$ are the $x$ and $y$ components 
of the vector potential, and the scalar potential $W_j(x,y) = \left[2 j \Omega _{\text{coh}}-j^2\gamma ^2 -2 j 
\gamma \Omega_{\rm rot} y -\Omega_{\rm rot} ^2 \left(x^2+y^2\right)\right]/2$. With these definitions, 
the effective potentials~\cite{radic} for $j = 0,\pm 1$ are given as
\begin{align}\label{Veff}
V_{\rm eff}^{j}(x,y) &= \frac{1}{2}\left[ (1-\Omega_{\rm rot}^2)(x^2+y^2)-j^2\gamma_x^2\right.\nonumber\\ 
&\left.+ 2 j\Omega_{\rm coh}-2 j \gamma_x\Omega_{\rm rot}y\right].
\end{align}
\begin{figure}[h]
\includegraphics[ width=0.49\textwidth]{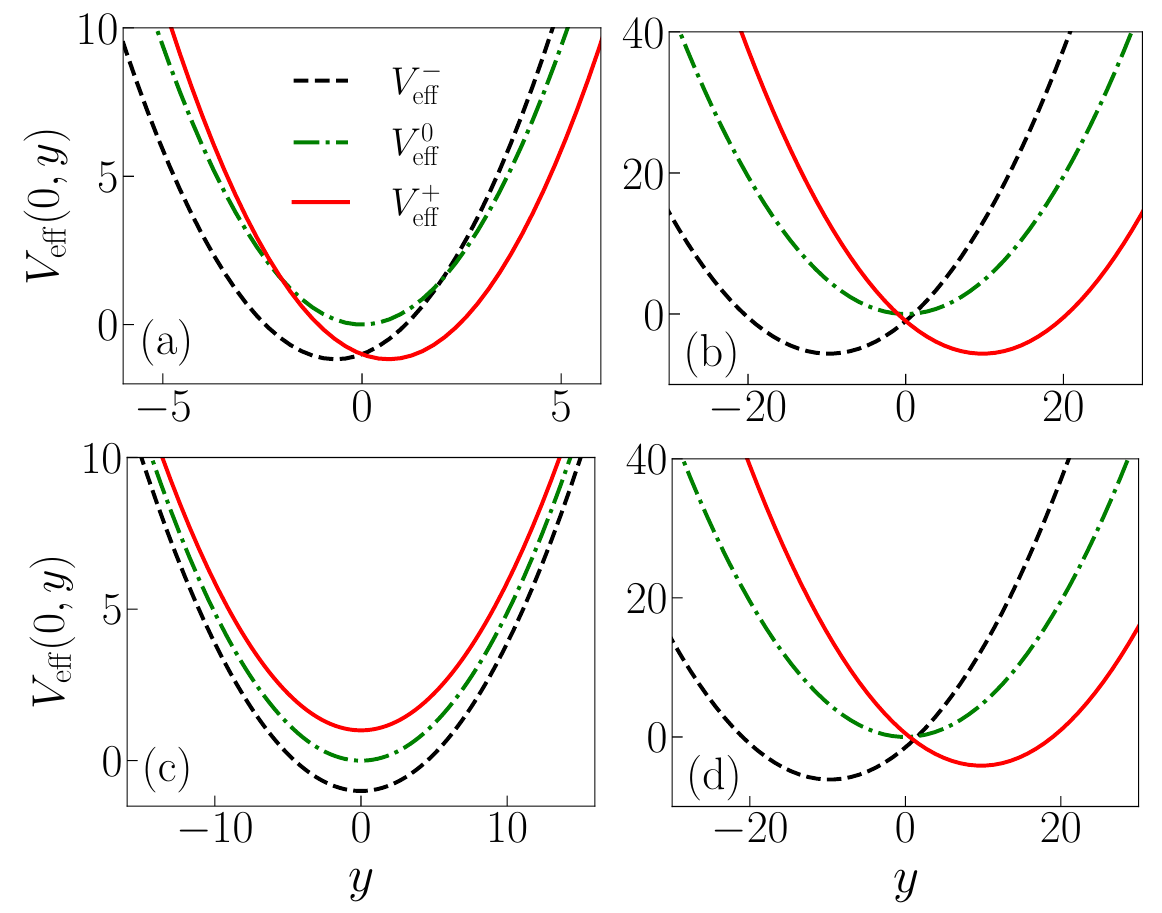}
\caption{(Color online) Sketch of the effective potential $V_{\rm eff}^{j}$, viz. Eq.~(\ref{Veff}), 
along $y$-axis: (a) $\gamma_x = 1,~\gamma_y=0,~\Omega_{\rm coh} = 0$, 
and $\Omega_{\rm rot} = 0.5$, (b) $\gamma_x = 1,~\gamma_y =0,~\Omega_{\rm coh} = 0$, 
and $\Omega_{\rm rot} = 0.95$, (c)  $\gamma_x = 0,~\gamma_y=0,~\Omega_{\rm coh} = 1$, 
and $\Omega_{\rm rot} = 0.95$, and (d) $\gamma_x = 1,~\gamma_y=0,~\Omega_{\rm coh} = 1$, 
and $\Omega_{\rm rot} = 0.95$.}
\label{effec+pot}
\end{figure}
From Eq.~(\ref{Veff}), $V_{\rm eff}^{+1}(x,y)$ and $V_{\rm eff}^{-1}(x,y)$ overlap at 
$y = \Omega_{\rm coh}/\gamma_x\Omega_{\rm rot}$ for $\gamma_x\ne 0$ and  $\Omega_{\rm rot}\ne 0$. 
In the region, $y<\Omega_{\rm coh}/\gamma_x\Omega_{\rm rot}$, $V_{\rm eff}^{-1}$ is lower than than the
other two and with a minima at $\gamma_x\Omega_{\rm rot}/(1-\Omega_{\rm rot}^2)$, whereas for
$y>\Omega_{\rm coh}/\gamma_x\Omega_{\rm rot}$, $V_{\rm eff}^{+1}$ is the low lying potential curve with 
a minima at $-\gamma_x\Omega_{\rm rot}/(1-\Omega_{\rm rot}^2)$. Which are also the positions of the 
density maxima of $|\Phi_{\pm 1}(x,y)|^2$, as discussed earlier. 
We illustrate the effective potential energy curves for the two representative cases
with $\gamma_x = 1$ and $\gamma_y = \Omega_{\rm coh} = 0$: under a moderate rotation frequency ($\Omega_{\rm rot} = 0.5$) in 
Fig.~\ref{effec+pot}(a) and a high rotation frequency $(\Omega_{\rm rot} = 0.95)$ in Fig.~\ref{effec+pot}(b) .
The potentials thus 
experienced by the boson are effectively equivalent to symmetric double-well potentials with minima 
occurring at $(x=0,y \mp 0.67)$ and $(x=0,y {\color{magenta}\mp 9.74)}$, respectively.
Depending on the values of $\gamma_x$ and $\gamma_y$, the presence of the coherent coupling 
modifies the effective potential landscape in different ways, for example, with $\Omega_{\rm coh} = 1$ 
it is harmonic potential for $\gamma_x=\gamma_y=0$ with minima at origin and an asymmetric double-well 
potential for $\gamma_x=1$ and $\gamma_y=0$ with a global minima at $(x=0, y = - 9.74)$. 
These are, respectively, shown in Figs.~\ref{effec+pot}(c) and (d).

For the latter case (\ref{case-b}), namely  $\gamma_x = \gamma_y =\gamma$ and $\Omega_{\rm coh } = 0$, 
the eigenvalue problem for the Hamiltonian $H_0$ is not exactly solvable. 
We, therefore, use the variational method
to calculate an approximate minimum-energy solution  
considering the following variational {\em ansatz} in polar coordinates
\begin{equation}
\begin{split}
\Phi_{\rm var}(r, \phi)=&\frac{\exp{ \left(-\frac{r^2}{2 \sigma ^2}\right)}}{\sqrt{\pi  \sigma ^{2 n+4} \Gamma 
(n+2)}}\times(i A_1 r^{|n|} e^{i n\phi },\\& -A_2 r^{|n+1|} e^{i (n+1)\phi }, 
i A_3 r^{|n+2|}e^{i(n+2)\phi})^T,\label{c2-ansatz}
\end{split}
\end{equation}
where $A_1,A_2,A_3$ are the variational amplitudes, $\sigma$ is the variational width of the 
{\em ansatz}, and $n$ is a variational integer. In the absence of rotation, the ground state of the single
particle Hamiltonian is a circularly symmetric $(-1,0,+1)$ type multi-ring solution with $\pm 1$ 
components hosting $\mp1$ phase-singularities \cite{Ruokokoski2012-rapidcommu}. This allows us to fix the integer $n>=-1$.  
The normalization condition imposes the constraint
\begin{equation}
\left[\frac{A_1^2 \Gamma (|n| +1)}{\sigma^{2(1-|n|
{+}n)}\Gamma (n+2)}\right]
+ \left[A_3^2(n+2)\sigma ^2+A_2^2\right]=1,\label{norm_con}
\end{equation}
on the variational parameters $A_1,A_2,A_3,n$, and $\sigma$. The variational energy in this case is
\begin{multline}
  E_{\rm var}(A_1,A_2,A_3,n,\sigma)= \frac{\sigma ^{-2 (n+2)}}{2\Gamma (n+2)} [(\sqrt{2} 
  A_1 A_2 \gamma \\(-| n| +3 n+2) \sigma ^{| n| +n+2} \Gamma \left( \frac{1}{2} (n+| n| +2)\right)
  +A_1^2 \sigma ^{2| n| } \Gamma (| n|+1)\\((\sigma ^4+1) | n| -2 n \sigma ^2 \Omega_{\rm rot} +\sigma ^4+1)+(n+1) 
  \sigma ^{2 n+2} \Gamma (n+1)\\ (-2 \sqrt{2} A_2 A_3 \gamma  (n+2) \sigma ^2+A_3^2 (n+2) 
  \sigma ^2(n (\sigma ^4-2 \sigma ^2 \Omega_{\rm rot} +1)+\\3 \sigma ^4-4 \sigma ^2 \Omega +3)
  +A_2^2(n \sigma ^4-2 n \sigma ^2 \Omega_{\rm rot} +n+2 \sigma ^4-2 \sigma ^2 \Omega_{\rm rot} +2))].\label{Evar_iso}
\end{multline}
This energy can be minimized with respect to all variational parameters subject to the constraint in Eq.
(\ref{norm_con}) to fix the variational parameters. To illustrate the validity of the variational method
in this case, we consider three sets of parameters
\begin{subequations}
\begin{align}
\quad& \gamma = 0.5,\quad \Omega_{\rm coh} = 0, \quad \Omega_{\rm rot} = 0.95,\label{SET_NEW}\\
\quad& \gamma = 1,\quad \Omega_{\rm coh} = 0,\quad \Omega_{\rm rot} = 0.5,\label{SET_5}\\    
\quad& \gamma = 1,\quad \Omega_{\rm coh} = 0,\quad \Omega_{\rm rot} = 0.95.\label{SET_6} 
\end{align}
\end{subequations}
\begin{figure}[h]
\begin{center}
\includegraphics[width=0.49\textwidth]{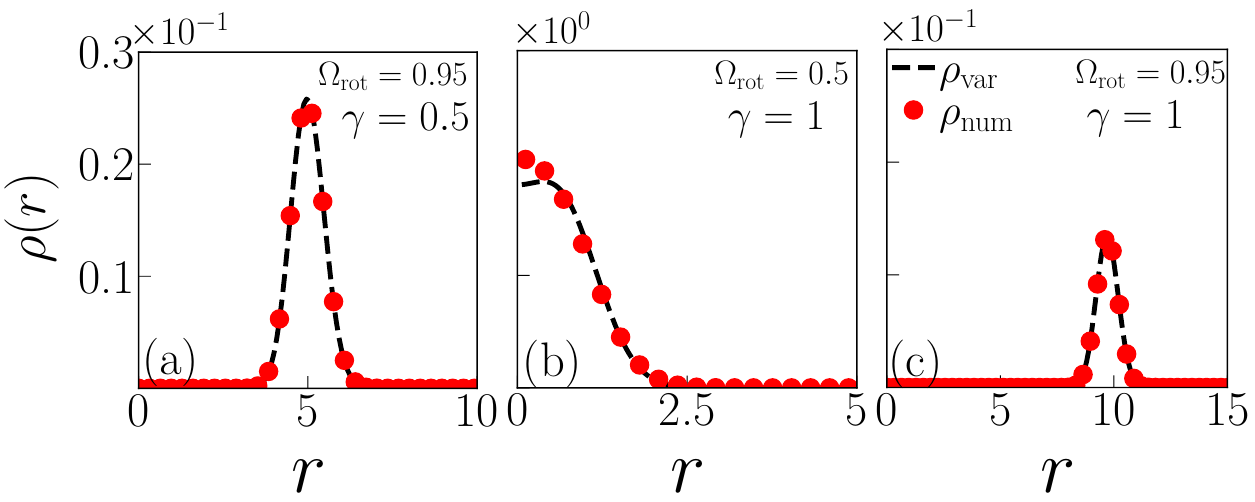}
\end{center}
\caption{(Color online) Total single-particle densities  corresponding to 
variational ($\rho_{\rm var}$) and exact numerical solution ($\rho_{\rm num}$) of 
the eigen-value problem for (a) $\gamma=0.5,~\Omega_{\rm coh}=0,~\Omega_{\rm rot}=0.95$; 
(b) $\gamma=1,~\Omega_{\rm coh}=0,~\Omega_{\rm rot}=0.5$; and (c) $\gamma=1,~\Omega_{\rm coh}=0,~\Omega_{\rm rot}=0.95$. 
The charges of phase singularities in the component wavefunctions corresponding to the total densities in (a), (b) and (c) are (+25,+26,+27), 
(0,+1,+2), and (+98,+99,+100), respectively.}
\label{compa_den2}
\end{figure}
The minimization of (\ref{Evar_iso}) results in $(A_1,A_2,A_3,n,\sigma) = (-2.512,~0.707,~0.097,~25,~0.975)$ for
parameters' set (\ref{SET_NEW}), $(A_1,A_2,A_3,n,\sigma) = (0.517,~-0.675,~-0.336,~0,~0.828 )$ 
for (\ref{SET_5}), and $(A_1,A_2,A_3,n,\sigma) = (-4.861,~ 0.707,~ 0.051,~98,~0.975)$ 
for (\ref{SET_6}). The comparison of
variational, $\rho_{\rm var}(r) = |\Phi_{\rm var}(r,\phi)^2|$,  and exact numerically evaluated single-particle
densitiy profiles, $\rho_{\rm num}(r)$, for (\ref{SET_NEW}), (\ref{SET_5}) and (\ref{SET_6}) are shown in 
Figs.~\ref{compa_den2} (a), (b) and (c), respectively.
The charges of phase singularities in the component wavefunctions obtained with the variational analysis, i.e., 
$(+25,+26,+27)$ for set (\ref{SET_NEW}), $(0,+1,+2)$ for set (\ref{SET_5}) and $(+98,+99,+100)$ for (\ref{SET_6})
match with the exact numerical results.
For the sets (\ref{SET_NEW}), (\ref{SET_5}), and  (\ref{SET_6}), the peaks of total 
variational densities lie along circles of radii $4.95$, $0.67$ and $9.74$, respectively, and
are reflective of the effective toroidal potential experienced by the boson.

\section{Rotating SO- and coherently-coupled spin-1 BEC}
\label{Section-III}
In a typical experiment the BEC can have atom-number ranging from a few thousands 
to up-to a few tens of a million, which primarily interact via $s$-wave scattering. 
At temperatures very close to absolute zero, this ultra-dilute quantum degenerate system is usually studied using 
a mean-field approximation which neglects the quantum and thermal fluctuations. In the mean-field approximation, 
a rotating SO-coupled spin-1 BEC in a quasi-two-dimensional harmonic trapping potential $V(x,y)$ can be described by three coupled 
GPEs \cite{plane-stripe,Ueda-review2012}, which in the dimensionless form are
 \begin{subequations}
 \begin{eqnarray}
i\frac{\partial \phi_{\pm1}}{\partial t}   &=&
 \mathcal{H}\phi_{\pm1} +{c_2}(\rho_0  \pm \rho_{-})\phi_{\pm1}+{c_2}\phi_{\mp1}^*\phi_0^2  \nonumber \\
  &&- \frac{i}{\sqrt{2}}\left(\gamma_x\partial_x\phi_0
 {\mp}i\gamma_y\partial_y\phi_0\right)+\frac{\Omega_{\rm coh}}{\sqrt{2}}\phi_0,\label{cgpet2d-1}\\
i\frac{\partial \phi_0}{\partial t} &=&
 \mathcal{H}\phi_0 +{c_2}\rho_{+}\phi_0+ 2{c_2}\phi_{+1}\phi_{-1}\phi_0^* -i\frac{\gamma_x}{\sqrt{2}}
 \nonumber\\ 
 &&\times\partial_x\left(\phi_{+1}+\phi_{-1}\right)
 +\frac{\gamma_y}{\sqrt{2}}\partial_y\left(\phi_{+1}-\phi_{-1}\right)\nonumber\\
 &&+\frac{\Omega_{\rm coh}}{\sqrt{2}}(\phi_{+1}+\phi_{-1}),\label{cgpet2d-2}
 \end{eqnarray}
\end{subequations}
\begin{align} 
\mathcal{H} &= \sum_{\nu=x,y}\frac{p^2_\nu}{2} +V(x,y)+{c_0}\rho - \Omega_{\rm rot} L_z\nonumber\\
\rho &= \sum_{j=\pm 1,0} \rho_j,~\rho_j = |\phi_j|^2,~\rho_{\pm}  = \rho_{+1}\pm\rho_{-1},\nonumber
\end{align}
where $c_0$ and $c_2$ are interaction parameters.
These
 are defined as
\begin{align}
c_0 = \sqrt{8\pi\alpha}&\frac{N(a_0+2a_2)}{3{a_{\rm osc}}},~
c_2 = \sqrt{8\pi\alpha}\frac{N(a_2-a_0)}{3{a_{\rm osc}}},
\label{interaction2d}
\end{align}
where $\alpha$ is the ratio of trapping frequency along axial direction to the radial $x$ direction, 
$N$ is the total number of atoms in the BEC, and $a_0$ and $a_2$ are the $s$-wave scattering 
lengths in total spin $0$ and $2$ channels, respectively. 
The coupled GPEs, viz. (\ref{cgpet2d-1})-(\ref{cgpet2d-2}), can be numerically solved using, for instance,
time-splitting methods  \cite{spin1-soc, spinf-soc,ravisankar-cpc,banger2021semiimplicit}. 


\subsection{Numerical solutions of coupled GPEs}
\label{Section-III-A}
We consider $10^5$ atoms of spin-1 BECs like $^{87}$Rb and $^{23}$Na
in an isotropic quasi-two-dimensional 
harmonic trap with $\alpha = 10$. The trapping frequencies are 
$\omega_x = \omega_y = 2\pi \times 10$ Hz resulting in $a_{\rm osc}^{\rm Rb} = 3.41~\mu$m and 
$a_{\rm osc}^{\rm Na} =6.63~\mu$m, respectively, for
$^{87}$Rb and $^{23}$Na spinor BECs. The ferromagnetic $^{87}$Rb has $a_0=101.8a_B$ and $a_2=101.4a_B$ 
\cite{scattering-Rb-spin1}, and anti-ferromagnetic $^{23}$Na has $a_0=50a_B$ and 
$a_2=55.01a_B$ \cite{scattering-Na-spin1}, where $a_B$ is the Bohr radius. The resultant dimensionless 
interaction strengths for $^{87}$Rb are $c_0 = 2482.21$ and $c_2 = -11.47$, and the same for $^{23}$Na are
$c_0=674.91$ and $c_2=21.12$. 
We solve coupled GPEs (\ref{cgpet2d-1})-(\ref{cgpet2d-2}) on a two-dimensional
512$\times$512 spatial grid with a spatial-step size $\Delta x= \Delta y = 0.1$ and a temporal step size $\Delta t = 0.005$
using a time-splitting Fourier spectral method \cite{spin1-soc}. We calculate the stationary-state
solutions by solving the coupled GPEs in imaginary time with an apt initial guess solution. 
In order to study the vortex-lattice states that
can emerge as the minimum energy solutions of an SO-coupled 
spin-1 BEC in a rotating frame, we consider the following SO-coupling strengths
\begin{equation}
\gamma_x = 1, \gamma_y=0;\quad \gamma = 0.5;\quad \gamma = 1,
\end{equation}
where as defined earlier $\gamma = \gamma_x = \gamma_y.$

\begin{figure}[!h]
\begin{center}
\includegraphics[width=0.46\textwidth]{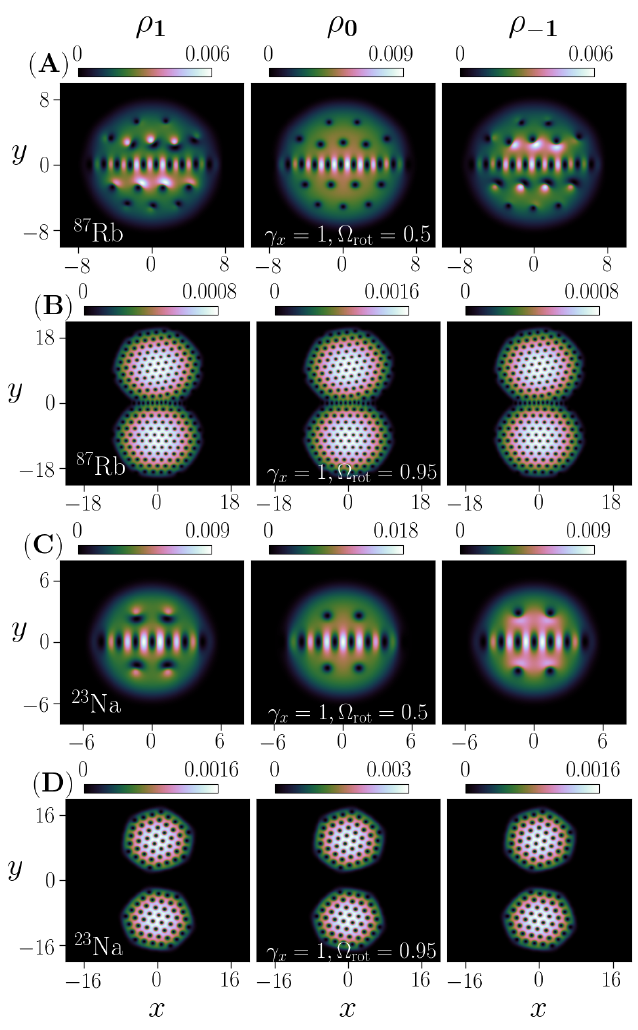}
\caption{(Color online) Equilibrium density profiles of the individual components of the SO-coupled $^{87}$Rb spin-1 BEC with
$c_0=2482.21$, $c_2=-11.47$, $\gamma_x=1$, and $\gamma_y=\Omega_{\rm coh}=0$: (A) with $\Omega_{\rm rot} =0.5$ and (B) with $\Omega_{\rm rot} =0.95$.
Similarly, (C) and (D) show the component densities for $^{23}$Na with
$c_0=674.91$ and $c_2=21.12$. The spatial coordinates and densities are in the units of $a_{\rm osc}$ and 
$a_{\rm osc}^{-2}$, respectively, where $a_{\rm osc} = 3.41$ $\mu$m for $^{87}$Rb and $6.63$ $\mu$m for $^{23}$Na.}
 \label{anti}
 \end{center}
 \end{figure}

We first study the rotating SO-coupled $^{87}$Rb and
$^{23}$Na spinor BECs with these SO-coupling strengths \emph{without} coherent coupling.
For which, we use the non-rotated ground states as the apt initial guess solutions to evolve the coupled GPEs
(\ref{cgpet2d-1})-(\ref{cgpet2d-2}) in imaginary time 
with $\Omega_{\rm rot}\ne 0$. For $\gamma_x = 1,~\gamma_y = 0$, the ground state is a plane-wave phase 
for $^{87}$Rb and a stripe phase for $^{23}$Na \cite{plane-stripe}. Here at small rotation frequencies, the phase-singularities (vortices) in the component wave functions 
exclusively align along the $x$-axis. The central-chain of holes in the individual component densities arising due to these 
phase-singularities with $\Omega_{\rm rot} = 0.5$ is evident in Fig.~\ref{anti}(A) for $^{87}$Rb and \ref{anti}(C)
for $^{23}$Na. 
  \begin{figure}[!h]
\begin{center}
\includegraphics[width=0.44\textwidth]{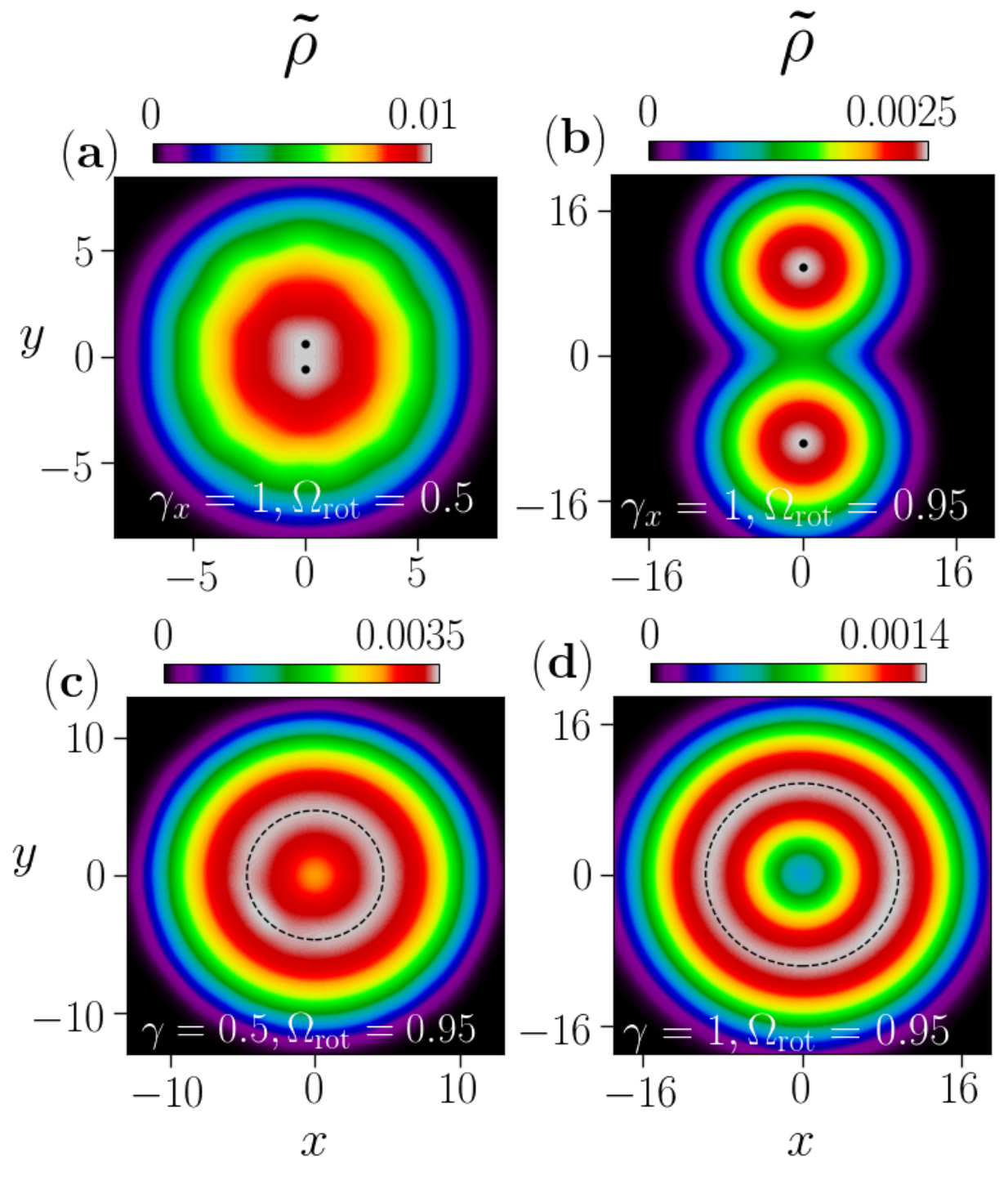}
 \caption{(Color online)
{(a) and (b), respectively, show the coarse-grained densities $\tilde{\rho}({\bf r})$ corresponding to
the total densities in Figs.~\ref{anti}(A) and (B); the peak values of
the coarse-grained densities at $(x = 0, y = \pm 0.6)$ and
$(x = 0, y = \pm 9.7)$ are marked by dots. Similarly,
(c) and (d), respectively, show $\tilde{\rho}({\bf r})$ corresponding to the solutions in Figs.~\ref{ferro}(B) and (D) 
and the respective peaks of $\tilde{\rho}({\bf r})$ are marked by dashed circles of radii $4.7$ and $9.7$.
The spatial coordinates and densities are in the units of $a_{\rm osc}$ and 
$a_{\rm osc}^{-2}$, respectively, where $a_{\rm osc} = 3.41~\mu$m.}}
 \label{coarseg}
 \end{center}
 \end{figure}
 \begin{figure}[!hbtp]
\begin{center}
\includegraphics[width=0.46\textwidth]{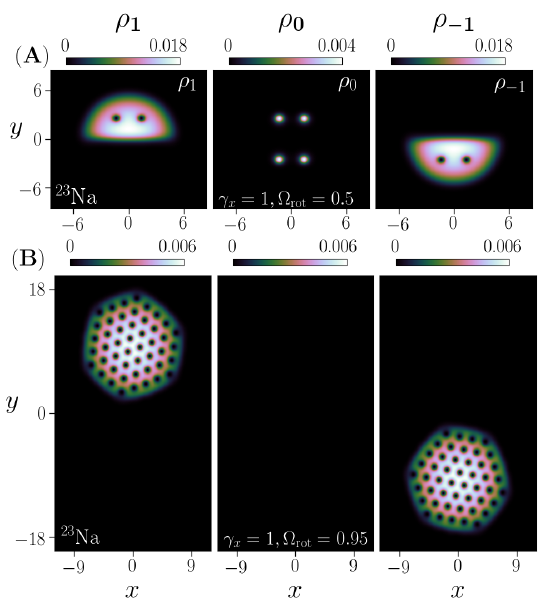}
\end{center}
\caption{(Color online) (A) and (B)
are the individual component densities of the stationary SO-coupled spin-1 BECs of
$^{23}$Na corresponding to $H_{\rm SOC} = \gamma_x p_x S_z$ for $\gamma_x = 1$
with $\Omega_{\rm rot} = 0.5$ and $\Omega_{\rm rot} = 0.95$, respectively.
The $m_j = 0$ component is fully absent in (B).
These solutions have been obtained by operating $U^{\dagger}$  on
the solutions corresponding to $H_{\rm SOC} = \gamma_x p_x S_x$ shown in 
Figs.~\ref{anti}(C) and  Figs.~\ref{anti}(D) for $^{23}$Na BEC.
The spatial coordinates and densities are in the units of $a_{\rm osc}$ and 
$a_{\rm osc}^{-2}$, respectively, where $a_{\rm osc} = 6.63~\mu$m for  $^{23}$Na.}
\label{model2}
\end{figure}
At still higher rotation frequency of $\Omega_{\rm rot}=0.95$, 
majority of vortices arrange themselves on both the sides of the central chain of vortices 
as shown in Figs.~\ref{anti}(B) and \ref{anti}(D) for $^{87}$Rb and $^{23}$Na, respectively.
The appearance of central chain of vortices, which appears along the line of the
intersection of $V_{\rm eff}^{+1}(x,y)$ and $V_{\rm eff}^{-1}(x,y)$, is a generic feature of
these systems with a sufficiently strong one-dimensional SO coupling \cite{liu3,rotating_spin1-aniso}.
In both the systems, the symmetric effective 
double-well potential leads to the condensate occupying the pairs of potential minima at $(x = 0, y = \pm 0.67)$ and
$(x = 0, y = \pm 9.74)$, respectively, when rotated with $\Omega_{\rm rot} = 0.5$ and $0.95$.
 This can be seen more vividly in the coarse-grained total density defined
as $\Tilde{\rho}({\bf r}) = \int C({\bf r} - {\bf r'})\rho({\bf r'})d{\bf r'}$, where 
$C({\bf r} - {\bf r'})$ is a normalized Gaussian with a width larger than the average 
inter-vortex separation. The coarse-grained total density peaks at the minima of the 
effective potentials. To illustrate, we refer the reader to  
$\Tilde{\rho}({\bf r})$ in Figs.~\ref{coarseg}(A) and (B) corresponding to the total density in
Figs.~\ref{anti}(A) and (B), respectively. The $\Tilde{\rho}({\bf r})$ peaks 
at $(x = 0, y = \pm 0.6)$ and $(x = 0, y = \pm 9.7)$, respectively, in the two cases.
The role of effective potential on the ground-state solution, say $(\phi_{+1},\phi_0,\phi_{-1})^T$, 
becomes obvious if one considers the unitary transformation
$(\psi_{+1},\psi_0,\psi_{-1})^T = U^\dagger(\phi_{+1},\phi_0,\phi_{-1})^T$, where component wave-function $\psi_j$ is subjected to
an effective potential $V_{\rm eff}^j(x,y)$ as discussed in Sec.~\ref{Section-II}. 
The component densities obtained by transforming the solutions shown in Figs.~\ref{anti}(C) 
and (D) for $^{23}$Na, for instance, are shown in Figs.~\ref{model2}(A) and (B), which as
discussed in the Sec.~\ref{Section-II} are the solutions corresponding to $\gamma_x S_z p_x$ coupling
in the mean-field model. The coarse-grained peak values of densities of $j = \pm 1$ components (which
are not shown here)
occur at $(0,\pm 0.6)$ and $(0,\pm 9.7)$ when rotated with $\Omega_{\rm rot} = 0.5$ and $0.95$ are in 
agreement with the 
effective potentials in Figs.~\ref{effec+pot}(a) and (b), respectively.
In the absence of rotation, $\gamma_x S_zp_x$ SO coupling favors miscibility of 
$j = \pm 1$ components for anti-ferromagnetic interactions, whereas it leads to phase-separation
if the coupling strength is above a critical value for ferromagnetic interactions \cite{phase_sep}.
In the presence of rotation, the effective potential can lead to the phase-separation
not only for a ferromagnetic $^{87}$Rb (not shown here) but also for an antiferromagnetic 
$^{23}$Na as is seen in the component density profiles in Figs. \ref{model2}(A) 
and (B) for $^{23}$Na. The $j = 0$ component occupies the cores of vortices in $j = \pm 1$ component at 
$\Omega_{\rm rot} = 0.5$ in Fig.~\ref{model2}(A).  With an increase in rotation frequency, number of 
atoms in $j = \pm 1$ components keep on increasing at the cost of atoms in $j =0$ component. 
Hence at larger rotation frequency of $\Omega_{\rm rot} = 0.95$ in Fig. \ref{model2}(B), 
there are no atoms in $j = 0$ component. Another consequence of the phase-separation 
is that the spin-expectation per particle (which is independent of rotation in spin space) 
tends to approach one for all the results shown in
Fig. \ref{model2} and consequently in Fig.~\ref{anti}. Thus $\gamma_x S_z p_x$ SO coupling provides a simpler description
of the results in Fig.~\ref{anti}.

\begin{figure*}[!htbp]
\begin{center}
\includegraphics[width=0.8\linewidth]{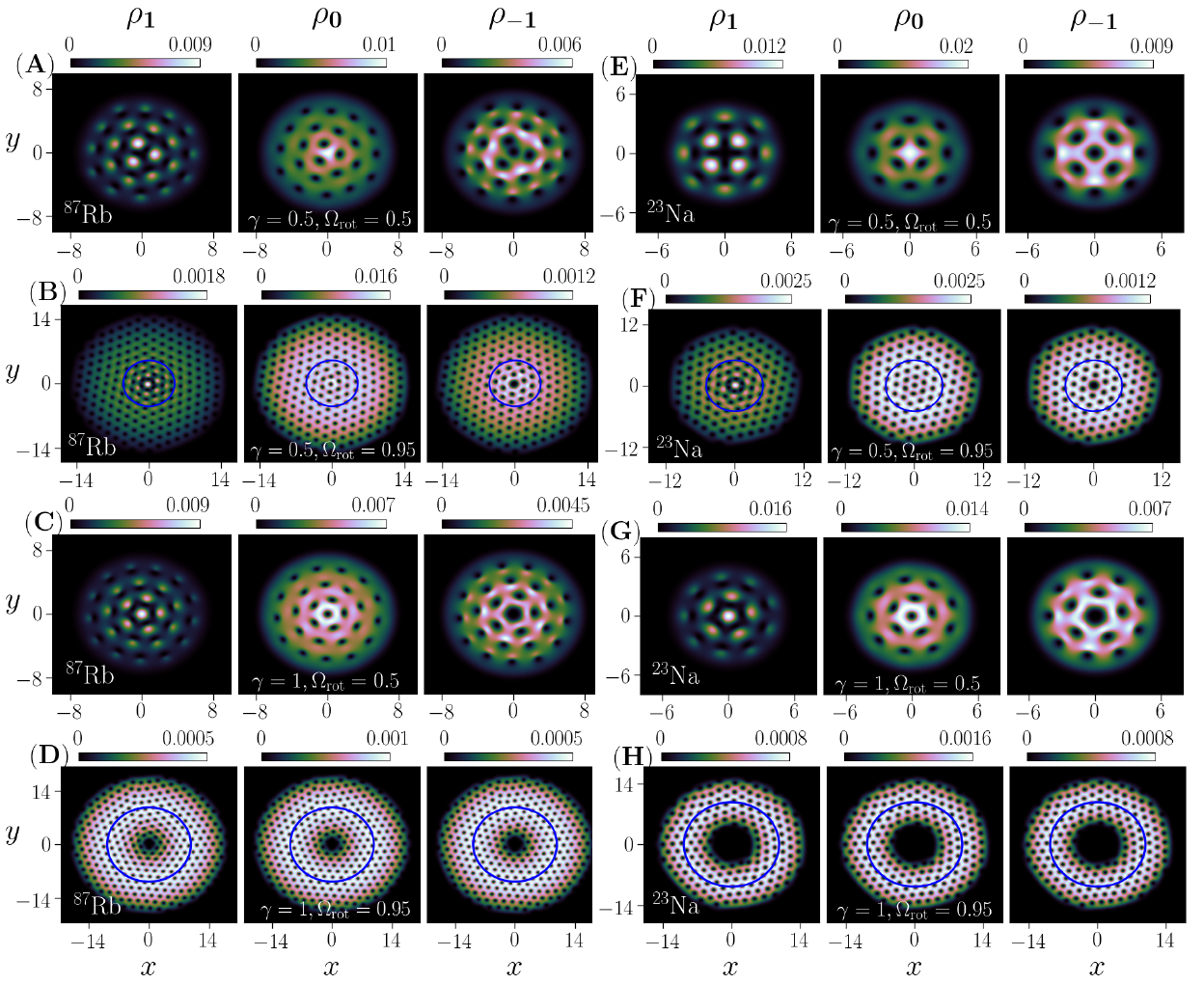}
\caption{(Color online) 
Equilibrium density profiles of the individual component densities of the SO-coupled spin-1 BECs:
(A)-(D) $^{87}$Rb with $c_0 = 2482.21$, $c_1 = -11.47$ and (E)-(H) $^{23}$Na spin-1 BEC with
$c_0=674.91$ and $c_2=21.12$. (A) and (B) have been obtained with 
$\Omega_{\rm rot} =0.5, 0.95$, respectively, and SO-coupling strength of $\gamma =0.5$. Similarly, 
(C) and (D) correspond to SO-coupling strength of $\gamma =1$ with
$\Omega_{\rm rot} = 0.5,$ and $0.95$, respectively. For SO-coupled $^{23}$Na, the plots (E) and (F) correspond to
$\gamma=0.5$ and $\Omega_{\rm rot}=$0.5,0.95, respectively, and (G) and (H)
correspond to $\gamma=1$ with $\Omega_{\rm rot}=0.5,0.95$, respectively. The spatial coordinates and densities are in the units of $a_{\rm osc}$ and 
$a_{\rm osc}^{-2}$, respectively, where $a_{\rm osc} = 3.41~\mu$m and $6.63~\mu$m for $^{87}$Rb and $^{23}$Na, 
respectively. }
\label{ferro}
\end{center}
\end{figure*}

Next for isotropic SO coupling  with $\gamma = 0.5$ 
and $\Omega_{\rm rot} = 0.5$,
the small number of vortices which nucleate are unable
to crystallize in a triangular vortex-lattice pattern as shown in Figs.~\ref{ferro}(A)
and \ref{ferro}(E) for $^{87}$Rb and $^{23}$Na, respectively, which are
consistent with the observations in Refs. \cite{vortex_lattice-adhikari,abrikosov}.
The vortex patterns in the component densities near the center in Fig.~\ref{ferro}(E)
resemble square-lattices consistent with a similar observation in Ref.~\cite{vortex_lattice-adhikari}.
The two condensates rotated at a high rotation frequency $\Omega_{\rm rot} = 0.95$ are shown in 
Figs.~\ref{ferro}(B) and \ref{ferro}(F); 
here the phase profiles of both the condensates (which are not shown) reveals that the center of the 
condensates have phase singularities of charges ($0,+1,+2$)
in $j = +1,~0$, and $-1$ components, respectively.
At this large rotation frequency more vortices are created in condensates which
relax in a triangular lattice pattern. The coarse-grained peak value of the total densities for the two condensates
lie along a circle of radius {$4.7$}, e.g. $\Tilde{\rho}({\bf r})$ corresponding to the solution
in Figs.~\ref{ferro}(B) is shown in Fig.~\ref{coarseg}(C), 
which {is in a decent agreement} with the variational single-particle {density 
maxima position in Fig.~\ref{compa_den2}(a)}. The circle encloses approximately $26$ phase singularities which agrees with 
phase-winding numbers calculated using variational analysis of the single-particle Hamiltonian. 
Next with $\gamma = 1$ and $\Omega_{\rm rot} = 0.5$, the component ground-state densities are shown
in Figs.~\ref{ferro}(C) and \ref{ferro}(G). The centers of both the condensates in this case 
have phase singularities of charges ($0,+1,+2$), respectively, in $j=+1,~0$, and $-1$ components, respectively. 
When rotated with higher frequency
of $\Omega_{\rm rot} = 0.95$, the condensate densities acquire a giant hole at the center as shown in 
Figs.~\ref{ferro}(D) and \ref{ferro}(H) with an annulus of triangular vortex-lattice pattern in each
component. The coarse-grained peak values of the total densities in this case too are along a
circle of radius $9.7$ as is seen in Fig.~\ref{coarseg}(D) for $^{87}$Rb, which agrees well with the variational single-particle density peak 
in Fig.~\ref{compa_den2}(c). The circle contains approximately $100$ phase sigularities in each component 
in agreement with the single particle. The appearance of a giant vortex at the trap center in 
the component densities surrounded by singly charged vortices arranged in an annulus for sufficiently 
strong isotropic SO-coupling strengths at fast rotations is a generic feature of these systems  
\cite{xu,aftalion}. The quantitative differences in respective component densities of $^{87}$Rb
and $^{23}$Na when rotated with $\Omega_{\rm rot} = 0.95$ is primarily a consequence of  
$c_0$ for the two BECs being $2482.21$ and $674.91$, respectively.

\subsubsection{Effect of Coherent coupling}
\begin{figure}[!h]
\begin{center}
\includegraphics[width=0.46\textwidth]{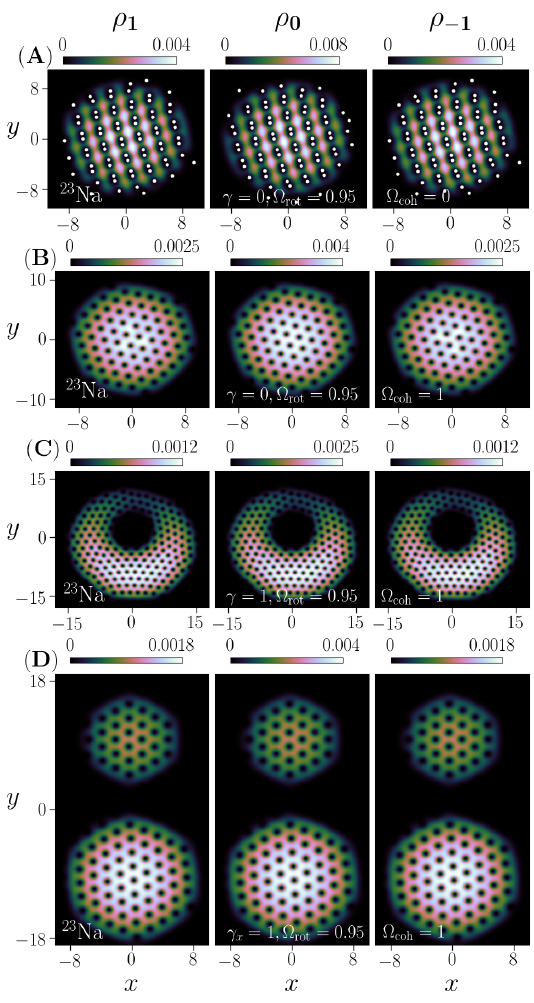}

\caption{(Color online) Equilibrium density profiles of the individual components of  $^{23}$Na spin-1 
BEC with interaction parameters $c_0=674.91$, $c_2=21.12$,
when rotated with $\Omega_{\rm rot} = 0.95$.
(A)
has been obtained for $\Omega_{\rm coh} = \gamma_x=\gamma_y=0$ and $\Omega_{\rm rot} =0.95$, whereas
(B), (C), and (D) have been obtained with coherent-coupling strength $\Omega_{\rm coh} = 1$ and 
SO-coupling strengths of $\gamma_x=\gamma_y=0$, $\gamma_x=\gamma_y=1$, and $\gamma_x=1,\gamma=0$, 
respectively. The spatial coordinates and densities are in the units of $a_{\rm osc}^{\rm Na}$ and 
$\left[a_{\rm osc}^{\rm Na}\right]^{-2}$ respectively, where $a_{\rm osc}^{\rm Na} = 6.63~\mu$m.}
\label{CCRESULT}
\end{center}
\end{figure}

To highlight the effects which can solely be attributed to an interplay of rotation, 
coherent coupling, and interactions, we first consider $^{23}$Na BEC without 
and with coherent coupling at rotation frequency of $\Omega_{\rm rot} = 0.95$ in the absence of SO 
coupling. Here without coherent coupling, the BEC supports an array of 
double-core vortices  \cite{kasamatsu2005vortices} in each component which arrange themselves in a square-lattice pattern as is shown 
in Figs.~\ref{CCRESULT}(A). Each double-core vortex core consists of two non-overlapping phase 
singularities of unit charge each which have been marked with white dots in Fig.~\ref{CCRESULT}(A). 
With coherent coupling of $\Omega_{\rm coh} =1 $, the system at the same rotation frequency of 
$\Omega_{\rm rot} = 0.95$ hosts a triangular-lattice pattern in each component as shown in 
Fig.~\ref{CCRESULT}(B), where a typical vortex core in each component consists of a single 
phase singularity. Here the effective potential is an isotropic harmonic potential $V_{\rm eff}^{-1}$
as shown in shown in Fig.~\ref{effec+pot}(c).

Next we consider the combined effect of SO and coherent couplings on the ground-state vortex
configurations. 
Here we consider two parameters sets- first with $\gamma_x=\gamma_y=1$, 
$\Omega_{\rm coh} = 1$, and second with  $\gamma_x = 1, \gamma_y = 0$, $\Omega_{\rm coh} = 1$.
In the former case, the ground state density has a hole whose center is shifted along $+y$ direction as 
shown in Fig.~\ref{CCRESULT}(C). In the latter, the component
densities distribute in two unequal triangular lattice patterns above and below $x$-axis as shown in 
Fig.~\ref{CCRESULT}(D), and with an increase in $\Omega_{\rm coh}$, 
the size of smaller triangular lattice pattern in the upper-half plane decreases further with a corresponding 
increase in the size of one in the lower-half plane. 
The  splitting of the component densities in two unequal parts can be attributed to the effective potential 
experienced by the system which is an asymmetric double-well potential created by $V_{\rm eff}^{-1}(x,y)$ and
$V_{\rm eff}^{+1}(x,y)$ with a global minima at $(x=0,y=-9.7)$ and a local minima at $(x=0,y=+9.7)$ 
as shown in Fig.~(\ref{effec+pot})(d). We obtain similar results for $^{87}$Rb spin-1 BEC
at $\Omega_{\rm rot} = 0.95$ which have not been shown here. 

\subsection{Spin-expectation per particle and spin-texture}
\label{Section-III-B}
\begin{figure}[h]
\begin{center}
\includegraphics[width=0.49\textwidth]{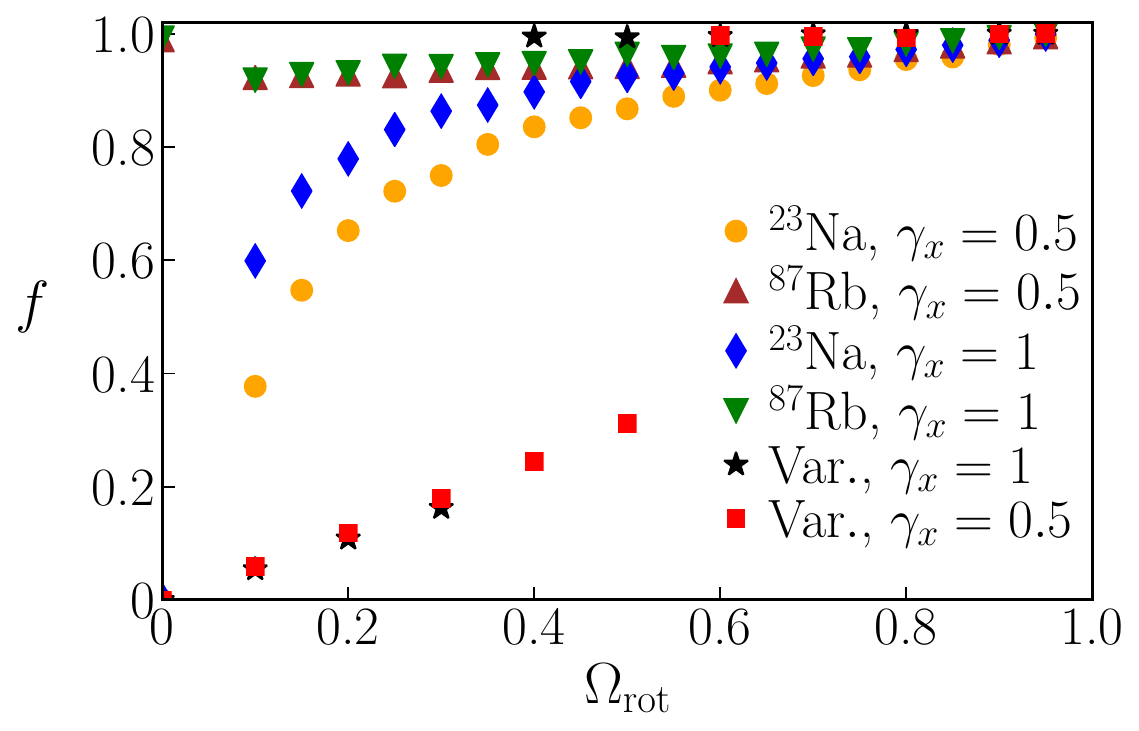}
\end{center}
\caption{(Color online) Variation of spin-expectation per particle $f$ with rotation frequency obtained from
the variational method discussed in Sec. \ref{Section-II} and the numerical solutions
of the GPEs for the SO-coupled $^{87}$Rb and $^{23}$Na BECs.}
\label{compa_spin_ang}
\end{figure}

\begin{figure}[!h]
\begin{center}

\includegraphics[width=0.41\textwidth]{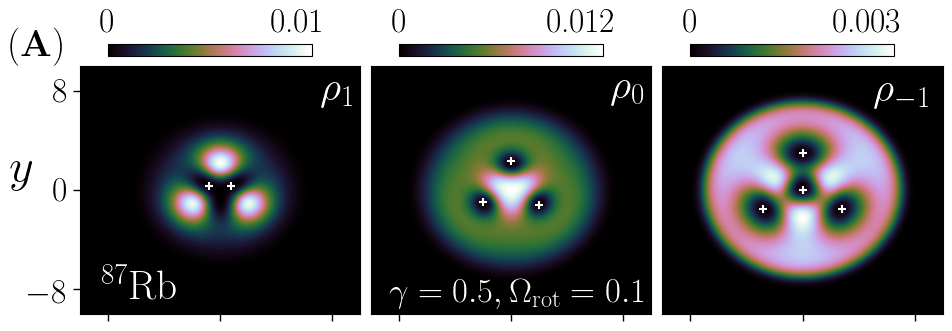}\\[0.01 cm]
\includegraphics[width=0.41\textwidth]{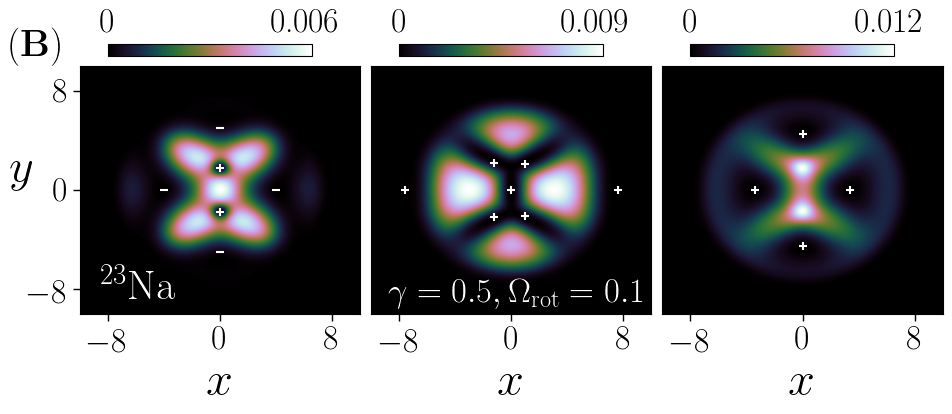}\\[0.01 cm]
\includegraphics[width=0.47\textwidth]{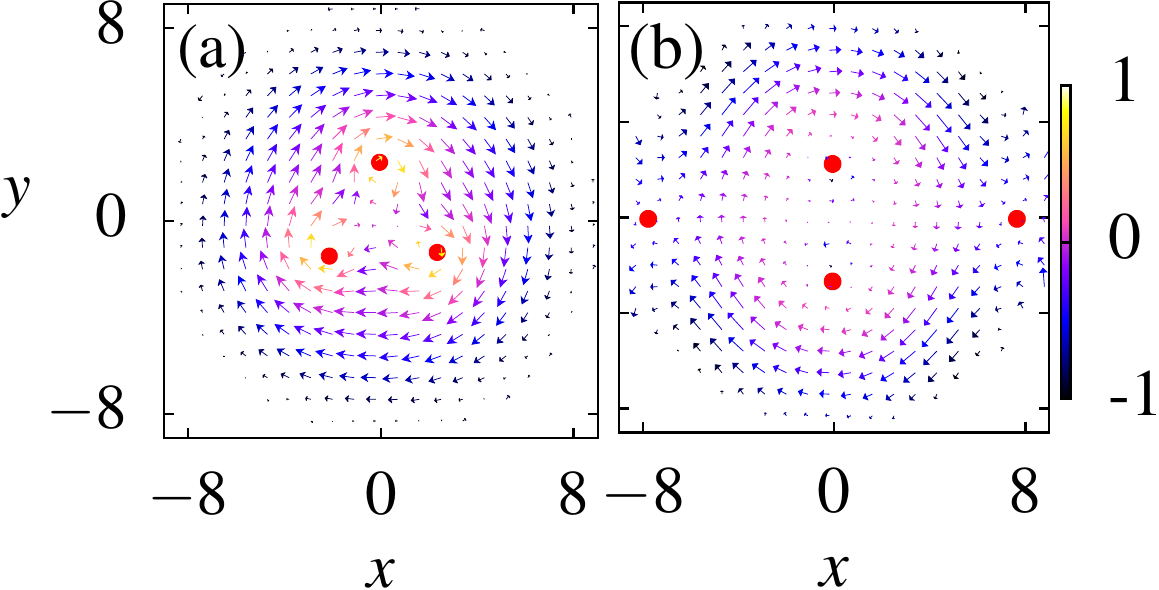}
\caption{(Color online) 
(A) displays the component density of the SO-coupled $^{87}$Rb 
BEC with $\gamma_x = \gamma_y = 0.5$ when rotated with $\Omega_{\rm rot} = 0.1$ and (B) displays the same for $^{23}$Na BEC. 
The interaction strengths for $^{87}$Rb and $^{23}$Na are $c_0=2482.21,~c_2=-11.47$ and 
$c_0=2482.35,~c_2=77.68$, respectively. The locations and signs of phase-singularities in each component are {marked} with $\pm$ signs.
(a) and (b), {respectively, show the spin-textures} corresponding to the {densities in} (A) and (B), where
(a) has the three skyrmions {(marked by red dots)}, and (b) has two near the center
of the trap in addition to two cross-disgyrations in 
spin-texture along {$x$-axis} coinciding
with $+1$ phase singularity in $m_f= 0$ component.  The spatial coordinates and densities are in the units of $a_{\rm osc}$ and 
$a_{\rm osc}^{-2}$, respectively, where $a_{\rm osc} = 3.41~\mu$m and $6.63~\mu$m for $^{87}$Rb and $^{23}$Na, 
respectively.
}
\label{texture0.1}
\end{center}
\end{figure}

\begin{figure}[h]
\begin{center}
\includegraphics[width=0.48\textwidth]{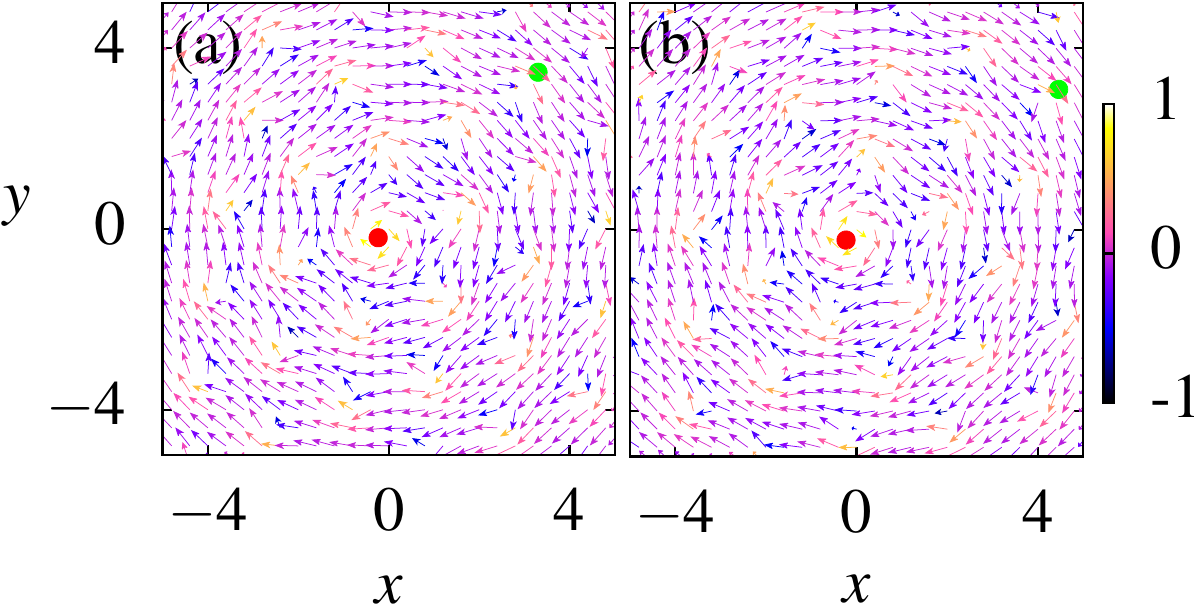}
\caption{(Color online) (a) shows the spin-texture for $^{87}$Rb system and (b) shows the same for $^{23}$Na
system at rotation frequency $\Omega_{\rm rot}=0.95$.
Both figures have a skyrmion at centre and in rest of the
regions half skyrmions lattice. The respective interaction parameters are same as those considered in Fig.~\ref{texture0.1}.}
\label{texture0.95}
\end{center}
\end{figure}

As noted in Sec.~\ref{Section-III-A}, the ground state solutions of the
rotating SO-coupled $^{87}$Rb and $^{23}$Na BECs at moderate
to high rotation frequencies are qualitatively similar, and
the quantitative differences stem from the different magnitudes of $c_0$.
To ascertain this further, here we
consider SO-coupled $^{87}$Rb and $^{23}$Na spin-1 BECs with  $\gamma=0.5~{\rm or}~ 1$,
$\Omega_{\rm coh}=0$ and (approximately) same $c_0$ but with different atom numbers. For $^{87}$Rb, 
we again consider $c_0 = 2482.21$ and $c_2 = -11.47$ corresponding to 10$^5$ atoms, whereas for 
$^{23}$Na we consider 3.68$\times$10$^5$ atoms resulting in $c_0 = 2482.35$ and $c_2 = 77.68$. We define the
spin-density vector ${\bf F} = (F_x,F_y,F_z)$ where
\begin{equation}
F_{\nu}(x,y) = \sum_{m,m'} \phi^*_m(x,y) (S_{\nu})_{m m'} \phi_{m'}(x,y),
\label{spintexture}
\end{equation}
and $f=\int |{\bf F}(x,y)|d{\bf r}/\int \rho(x,y) d{\bf r}$, which serves as a measure of spin-expectation 
per particle for an inhomogeneous system.  We examine the angular momentum per particle, $f$, and spin-texture 
\cite{Ueda-review2012} ${\bf f}(x,y) = {\bf F}(x,y)/\rho(x,y)$ as a function of rotation frequency. 
In the absence of rotation, the $^{87}$Rb and $^{23}$Na spin-1 BECs have $f = 1$ and $0$, respectively \cite{Ueda-review2012}.
The $f$ as a function of rotation frequency $\Omega_{\rm rot}$ for the two systems is 
shown in Fig~\ref{compa_spin_ang}, which illustrates that with increase in $\Omega_{\rm rot}$, 
$f\to 1$ for $^{23}$Na whereas it remains close to $1$ for $^{87}$Rb. 
We also analyse spin-expectation per particle using the single-particle variational solution $\Phi_{\rm var}$ 
in Eq.(\ref{c2-ansatz}) to evaluate $f$. The variational analysis predicts $f\approx 1$ for 
$\gamma = 0.5~(1)$ and $\Omega_{\rm rot}\ge 0.6~(0.4)$, 
which is consistent with the numerical results for $^{87}$Rb and $^{23}$Na BECs at moderate to
high rotations as is shown in Fig.~\ref{compa_spin_ang}. 
The differences in numerical and variational $f$ values for $\Omega_{\rm rot} \le0.6~(0.4)$ are mainly 
because of  spin-dependent interactions, which expectedly become increasingly less important with an increase in rotation
frequency.
Next we consider the spin-texture of
$^{87}$Rb and $^{23}$Na BEC with $\gamma=0.5, \Omega_{\rm coh} = 0$ when rotated with
$\Omega_{\rm rot} = 0.1$ and $0.95$. 
{The component densities 
for $^{87}$Rb and $^{23}$Na when rotated with $\Omega_{\rm rot} = 0.1$ are shown in Figs.~\ref{texture0.1}(A) 
and (B), respectively, and the corresponding spin textures are in 
Figs.~\ref{texture0.1}(a) and (b).}
At this frequency $^{87}$Rb hosts three skyrmions in {Fig.~\ref{texture0.1}(a)
as compared to two for $^{23}$Na
 in Fig.~\ref{texture0.1}(b)}
(near the center of the trap). The generation of skyrmion and half-skyrmion
excitations in rotating SO-coupled BECs is discussed in Refs.~\cite{liu1,liu2}. The spin-textures at 
$\Omega_{\rm rot} = 0.95$ are shown in Fig.~\ref{texture0.95}{(a)} for $^{87}$Rb and Fig.~\ref{texture0.95}{(b)} 
for $^{23}$Na, here both the systems have a skyrmion at the center surrounded by a lattice of half-skyrmions.
The spin-texture of $^{87}$Rb corresponds to the component densities shown
in Fig.~\ref{ferro}(B), 
whereas the component densities of $^{23}$Na which are indistinguishable $^{87}$Rb are not shown here.
The similarity of the two systems at faster rotation is therefor also reflected in the spin-textures.
The similarity in the response of the two systems at fast rotations has also been confirmed based upon
their mass and spin-currents.

\section{Summary and conclusions}
\label{Section-IV}
We have studied the effects of SO- and coherent-coupling on rotating ferro- and antiferromagnetic spin-1 BECs. 
Using exact numerical solutions complemented by a variational analysis, we have shown that the non-interacting 
part of the underlying spin-1 Hamiltonian can be translated to the rotating effective potentials with symmetric, 
asymmetric double-well, and toroidal structures. To this end, we have illustrated that using the mean-field 
Gross-Pitaevskii formalism, employing the realistic experimental parameters, the spatial distribution of $^{87}$Rb and 
$^{23}$Na BECs are consistent with the inhomogeneity of the effective potentials. The effects of rotation are 
further elucidated by computing the spin-expectation per particle for the ferro- as well as the antiferromagnetic 
BECs. For the former, the spin-expectation is always close to unity irrespective of the rotation frequency. 
While, for the latter, the spin-expectation value increases with an increase in rotation frequency 
and tends to approach one. For the simpler one-dimensional coupling ($\propto \gamma S_z p_x$), spatial segregation 
between the $j = \pm 1$ components results in spin-expectation per particle approaching one for the antiferromagnetic 
BEC; similarly, single-particle variational analysis with Rashba SO coupling also indicates the spin-expectation 
per particle approaching one irrespective of the spin-exchange interactions with increasing rotational frequency. 
The similarity in response of the fast-rotating ferromagnetic $^{87}$Rb and antiferromagnetic $^{23}$Na 
highlights the much diminished role of the spin-exchange interactions vis-{\`a}-vis the other competing terms
in the system's Hamiltonian.

\section*{Acknowledgements}
AR acknowledges the support of Science \& Engineering Research Board (SERB), Department of Science and Technology, 
Government of India under the project SRG/2022/000057 and IIT Mandi seed-grant funds under the project IITM/SG/AR/87.
RKK is supported by the Marsden Fund of New Zealand (Contract No. UOO1726).
SG acknowledges support from the Science and Engineering Research Board, Department of Science and
Technology, Government of India through Projects No.
ECR/2017/001436 and No. CRG/2021/002597.
\end{document}